\documentclass[onecolum,secnumarabic,amssymb, nobibnotes,longbibliography,preprint]{revtex4-1}

\usepackage{graphicx,color,amsmath}
\usepackage[normalem]{ulem}

\begin{document}

	\title{A First-Quantized Model For Unparticles and Gauge Theories Around Conformal Window}%

	\author{N. Boulanger$^1$}
	\email[e-mail:]{nicolas.boulanger@umons.ac.be}
	\author{F. Buisseret$^{2,3}$}
	\email[e-mail:]{buisseretf@helha.be}
	\author{G. Lhost$^{1}$}

	\affiliation{$^1$ Service de Physique de l'Univers, Champs et Gravitation, Universit\'{e} de Mons, UMONS  Research Institute for Complex Systems, Place du Parc 20, 7000 Mons, Belgium }
	\affiliation{$^2$ CeREF, Chaussée de Binche 159, 7000 Mons, Belgium}
	\affiliation{$^3$ Service de Physique Nucl\'{e}aire et Subnucl\'{e}aire, Universit\'{e} de Mons, UMONS Research Institute for Complex Systems, 20 Place du Parc, 7000 Mons, Belgium}

	\date{\today}%
	
	\begin{abstract}
	We first quantize the action proposed by Casalbuoni and Gomis 
	in [Phys. Rev. D \textbf{90}, 026001 (2014)], 
	an action that describes two massless relativistic scalar particles interacting 
	via a conformally invariant potential. 
	The spectrum is a continuum of massive states that 
	may be interpreted as unparticles. 
	We then obtain in a similar way the mass operator for a deformed action 
	in which two terms are introduced that break the conformal symmetry:
	a mass term and an extra position-dependent coupling constant. 
	A simple Ansatz for the latter leads to a mass operator with linear confinement 
	in terms of an effective string tension $\sigma\,$. 
	The quantized model is confining when $\sigma\neq0$ and its mass spectrum 
	shows Regge trajectories. We propose a tensionless limit in 
	which highly excited confined states reduce to (gapped) unparticles. 
	Moreover, the low-lying confined bound states become massless 
	in the latter limit as a sign of conformal symmetry restoration 
	and the ratio between their masses and $\sqrt\sigma$ stays constant. 
	The originality of our approach is that it applies to both confining and
	conformal phases via an effective interacting model.
	\end{abstract}
	
	\maketitle
	
	\section{Introduction}
	
	It is known that some asymptotically free gauge theories with $N_f$ light 
	fermion flavours and $SU(N)$ gauge group have a conformal window, i.e., 
	there exists an energy range in which the beta function vanishes. 
	This conformal window occurs at various values of $N_f$ and $N\,$, 
	depending on the fermion representation, see \cite{Dietrich:2006cm} 
	for an extensive list of examples obtained by inspection of 
	the two-loop beta function. For example, $SU(3)$ gauge theory with 
	12 light quarks flavours in the fundamental representation is 
	conformal, as confirmed by a five-loop and nonperturbative calculations 
	\cite{DiPietro:2020jne,Hasenfratz:2016dou}. 
	$SU(2)$ gauge theory with 2 quark flavours in the adjoint representation 
	can also be mentioned \cite{Hietanen:2008mr,DelDebbio:2015byq}, 
	or 2 quark flavours in the two-index symmetric 
	representation \cite{Catterall:2007yx}. 
	Note that conformal windows are expected to be present for other gauge 
	groups than $SU(N)$, like $Sp(2N)$ and $SO(N)$ \cite{Sannino:2009aw}. 
	Many evidences showing the existence of a conformal window for specific 
	gauge theories have been found by 
	resorting to lattice QCD methods, see e.g. the reviews 
	\cite{Neil:2012cb,Witzel:2019jbe,USQCD:2019hee,Drach:2020qpj,Cacciapaglia:2020kgq}. 
	General algorithms actually exist for 
	$SU(N)$ theories with quarks in arbitrary representations \cite{DelDebbio:2008zf}.
	
	To our knowledge, no effective model -- we mean a simple enough action so as 
	to allow for analytical calculations -- has been proposed to mimic confining 
	gauge theories 
	when they approach the conformal window starting from a confining phase. 
	Our starting point has been proposed in Ref. \cite{Casalbuoni:2014ofa}, 
	where an action describing two scalar relativistic particles with 
	conformal invariant interactions has been presented. The latter action reads
	\begin{equation}\label{S0}
		S_C = \alpha\int d\tau\left(\frac{\dot x^2_1\dot x^2_2}{r^4}\right)^{1/4}\;,
	\end{equation}
	with $\alpha$ a dimensionless parameter, 
	$x^\mu_i=x^\mu_i(\tau)$ the parametric equations for the two particles in 
	the $(D+1)$-dimensional Minkowski spacetime with metric 
	$\eta=$ diag$(-+\dots+)\,$ in inertial coordinates, 
	$\tau$ an evolution parameter and $r^\mu:=x^\mu_2-x^\mu_1\,$. 
	For any vector with components $v^\mu\,$, 
	$v^2:=\eta_{\mu\nu}v^\mu v^\nu =: v\cdot v$  is the squared
	norm in the Lorentzian sense. 
	In Section \ref{sec:review} we first review and quantize the 
	action $S_C$. In particular, while reproducing 
	some steps of the canonical analysis of \cite{Casalbuoni:2014ofa}, we also 
	detail some issues that were not presented therein for the sake of conciseness 
	but that we need for the quantization of the model. 
	We link the spectrum obtained after quantization to unparticles, 
	originally introduced as a nontrivial scale-invariant sector in 
	low-energy effective field theories \cite{Banks:1981nn},  
	see also \cite{Georgi:2007ek,Georgi:2008pq}. 
	
	The model is generalized in two ways that break conformal invariance. In
	Sec. \ref{sec:massive} we introduce a mass term for the interacting particles. 
	The quantization leads to unparticle spectrum with a mass gap. 
	Then a confining interaction is introduced in Sec. \ref{sec:model} 
	via a change of the form $\alpha^2\to \alpha^2U^2(r^2)\,$. 
	The quantization of our model is then performed and the spectrum is 
	analytically computed. The transition from the confining phase to the 
	conformal phase is finally studied and unparticles are shown to emerge 
	from the confined spectrum. The idea that unparticles 
	may be "hidden" in a gauge theory's conformal window has been developed 
	in \cite{Ryttov:2007sr}. 
	We finally argue that the present model provides a toy model to 
	illustrate the latter proposal.  

	\section{Conformal phase}\label{sec:review}
	\subsection{Classical analysis}
	Let us review the conformally-invariant action presented by Casalbuoni and Gomis in Refs. 
	\cite{Casalbuoni:2014ofa,Casalbuoni:2014qia}. 
	The aim of their presentation was to show the existence of higher-spin-type 
	conserved currents in their model while ours is rather to prepare 
	the ground for the quantization of our model.
	
	In the case of two massless particles, 
	the authors of \cite{Casalbuoni:2014ofa} start from the action
	\begin{equation}
		S_0[x_1^\mu,e_1,x_2^\mu,e_2] = \int d\tau \;
		\Big( \frac{\dot{x}_1^2}{2e_1} + \frac{\dot{x}_2^2}{2e_2} \Big)\;, 
	\end{equation}
	with $e_i$ $(i=1,2)$ the einbeins, 
	and add an interaction term that couples the two 
	massless particles in a conformally-invariant way:
	\begin{equation}
		S[x_1^\mu,e_1,x_2^\mu,e_2] = \int d\tau\; \Big(    \frac{\dot{x}_1^2}{2e_1} + 
		\frac{\dot{x}_2^2}{2e_2}  - \frac{\alpha^2}{4} \frac{\sqrt{e_1 e_2}}{r^2}   
		\Big)\;,
		\label{actioninteraction1}
	\end{equation}
	where we recall that $r^2:=r^\mu r^\nu\,\eta_{\mu\nu}=: r\cdot r$
	and where we have chosen the potential to be repulsive. 
	In the above action, the coupling 
	constant $\alpha$ is dimensionless. 
	Since the action is manifestly Poincar\'e invariant it suffices 
	to prove that it is invariant under dilations and special conformal 
	transformations. As for the dilations, it is easy to see that 
	the transformations 
	$x_i^\mu \mapsto \lambda x_i^\mu\;$, $e_i \mapsto \lambda^2 e_i\,$
	preserve the action. 
	The invariance under special conformal transformations is established 
	by defining them as a succession of a Poincar\'e translation followed by 
	an inversion and another Poincar\'e translation, 
	where the inversion is the transformation
	\begin{equation}
		x_i^\mu ~\mapsto~ \frac{x_i^\mu}{x_i^2} \;,\qquad 
		e_i ~\mapsto~  \frac{e_i}{x_i^4}\;, \qquad 
		r^2 ~\mapsto~ \frac{r^2}{x_1^2 x_2^2}\;.
		\label{inversion}
	\end{equation}
	The interested reader may find additional information in the following 
	\href{https://www.damtp.cam.ac.uk/user/ho/CFTNotes.pdf}{lecture notes}. 
	The conformal invariance of the action is then readily checked. 
	
	Note that one can eliminate the two auxiliary variables $e_1(\tau)$ and $e_2(\tau)$
	by virtue of their own field equations, 
	which results \cite{Casalbuoni:2014ofa} in the incarnation (\ref{S0}) of the action. 
	We mention for completeness Ref. \cite{Pramanik:2014zfa} in which it is shown that 
	action (\ref{actioninteraction1}) may be modified in a simple way so that it models 
	two particles interacting conformally in Snyder space.
	
	%%%%%%%%%%%%%%%%%%%%%%%%%%%%%%
	The action (\ref{actioninteraction1}) does not depend on the derivatives 
	$\dot{e}_i\, $, hence there are two primary constraints:
	\begin{equation}
		\label{primaryC}
		\pi_1\approx 0\;, \qquad \pi_2\approx 0\;,
	\end{equation}
	where $\pi_i$ is the conjugate variable to $e_i\,$ and 
	where the symbol "$\approx$" denotes a weak equality, i.e., 
	an equality which holds on the constraint surface.
	One then derives the canonical Hamiltonian associated with 
	the Lagrangian action \eqref{actioninteraction1}:
	\begin{align}
		H_c &= \frac{e_1 p_1^2}{2} + \frac{e_2 p_2^2}{2}  + \frac{\alpha^2}{4} 
		\frac{\sqrt{e_1 e_2}}{r^2} + \dot{e}_1 \pi_1 + \dot{e}_2 \pi_2 
		\approx \frac{e_1 p_1^2}{2} + \frac{e_2 p_2^2}{2}  + \frac{\alpha^2}{4} 
		\frac{\sqrt{e_1 e_2}}{r^2}
		\label{Hc1}
	\end{align} 
	as well as the total Hamiltonian
	\begin{align}
		H_T(p_i,x_i,e_i,\pi_i, \lambda_1, \lambda_2) &= 
		\frac{e_1 p_1^2}{2} + \frac{e_2 p_2^2}{2}  
		+ \frac{\alpha^2}{4} \frac{\sqrt{e_1 e_2}}{r^2} 
		+ \lambda^1 \pi_1 + \lambda^2 \pi_2\;.
		\label{HT1}
	\end{align}
	The invariance of the primary constraint under the dynamical evolution 
	leads to two secondary constraints:
	\begin{align}
		\label{secondaryC}
		\lbrace  \pi_1, H_T  \rbrace = \frac{1}{2} \Big( -p_1^2 
		- \frac{\alpha^2}{4r^2} 
		\sqrt{\frac{e_2}{e_1}} \Big) =: \phi_1 \approx 0 \;,\qquad  &
		\lbrace  \pi_2, H_T  \rbrace = \frac{1}{2} 
		\Big( -p_2^2 - \frac{\alpha^2}{4r^2} 
		\sqrt{\frac{e_1}{e_2}} \Big) =: \phi_2 \approx 0 \;.
	\end{align}
	Pursuing the consistency algorithm with these two constraints gives 

	\begin{align}
		\lbrace  \phi_1 , H_T \rbrace &= -\frac{\alpha^2}{16 r^2}\, \Big( -\lambda_1 
		\sqrt{\frac{e_2}{e_1^3}}\, +  \frac{\lambda_2}{\sqrt{e_1 e_2}}\; \Big)  
		+ \frac{\alpha^2}{4r^4} \Big( \sqrt{e_1 e_2} \; p_1 \cdot r 
		+ \sqrt{\frac{e_2^3}{e_1}} \, p_2\cdot r \Big) =: \phi_3
		\label{phi3}
	\end{align}
	and $\{ \phi_2 , H_T \} = - \frac{e_1}{e_2}\,\phi_3\,$.
	One can identically solve $\phi_3 = 0$ by fixing one 
	of the Lagrange multipliers, say $\lambda_1(\tau)\,$:
	\begin{equation}
		\lambda_1 = \tilde{\lambda}_2 \,e_1 + C\;,
		\qquad \tilde{\lambda}_2 := \frac{\lambda_2}{e_2}\;,
		\qquad C := -\frac{4e_1}{r^2}\, ( e_1 \,p_1 \cdot r + e_2 \, p_2 \cdot r)\;,
		\label{lambda1}
	\end{equation}
	which gives rise to the following expression for the total Hamiltonian:
	\begin{align}
		& H_T = H_c + \lambda_2 \frac{e_1}{e_2} \pi_1 - 
		\frac{4e_1}{r^2} \,\big[ e_1 \,p_1 . r + e_2 \,p_2 . r \big] \pi_1 
		+ \lambda_2 \pi_2 
		= H_c + \tilde{\lambda}_2 (e_1\, \pi_1 + e_2 \,\pi_2 ) + C \,\pi_1\;.
		\label{totalH}
	\end{align}
	At this stage it is worth saying that the same analysis can be performed even if $\alpha$ is a function $\alpha(r)$: Although breaking conformal symmetry, this case is important for our purpose and will be discussed in Sec. \ref{sec:model}.
	
	Coming back to the Hamiltonian \eqref{totalH},
	since $\tilde{\lambda}_2$ is an arbitrary function, 
	one may take $e_1 \pi_1 + e_2 \pi_2$ as our new primary first-class constraint. 
	Its Poisson brackets with $\pi_1$ and $\pi_2$ is weakly zero 
	while it is strongly zero with $\phi_1$ and $\phi_2\,$. 
	Furthermore, $H_T$ is first-class since it does not explicitly depend on time.
	From the fact that the bracket of two first-class functions is first-class
	and the computation
	\begin{equation}
		\lbrace e_1 \pi_1 + e_2 \pi_2 , H_T \rbrace 
		= e_1 \,\phi_1 + e_2\, \phi_2 - C \,\pi_1 =:\gamma_2 \;, 
		\label{FC2}
	\end{equation}
	one derives that $\gamma_2$ is another first-class constraint. 
	To summarise, there are two first-class (FC) constraints 
	(one is primary and the other is secondary) 
	and two second-class (SC) constraints (one is primary and the other is secondary):
	\begin{align}
		\text{FC} &\quad:\quad \gamma_1 := e_1 \pi_1 + e_2 \pi_2 \quad\text{(primary)}\;, 
		&\gamma_2 := e_1 \phi_1 + e_2 \phi_2 - C \pi_1 \quad\text{(secondary)}\;,
		\label{contraintesfinales1}
		\\
		\text{SC} &\quad: \quad \chi_1 := \pi_1 \quad\text{(primary)}\;, 
		&\chi_2 := \phi_1 \quad\text{(secondary)} \;. 
		\label{contraintesfinales2}
	\end{align}
	One has
	\begin{equation}
		\lbrace \gamma_1 , \gamma_2 \rbrace = -\gamma_2\;,\qquad 
		\lbrace \gamma_2 , H_T \rbrace = \tilde{\lambda}_2\, \gamma_2\;,
		\qquad \lbrace \gamma_1 , H_T \rbrace = \gamma_2\;.
	\end{equation}

	The number $f$ of degrees of freedom is given by the 
	number of phase-space variables minus twice the number 
	of first-class constraints minus the number of second-class 
	constraints. The number of degrees of freedom for the conformally-invariant 
	interacting system \eqref{actioninteraction1} is therefore given by 
	$f = 4(D+1)+4 - 4 - 2 = 4D+2\,$, which differs from the counting obtained 
	for the non-interacting system that produces $f=4D\,$. 
	If one considers the free limit 
	$\alpha\rightarrow 0$, the limiting value should therefore not be considered. 
	This is also clear from the form \eqref{S0} of the action. 
	In fact, one can show that the Dirac conjecture is not satisfied by 
	the constrained system at hand. 
	It turns out that first-class constraints are all gauge symmetry generators 
	if four conditions established in Chapt. 3 of \cite{henneaux} are respected. 
	One of these conditions is that the secondary second-class constraints 
	should not appear in the Poisson bracket of the first-class constraints
	with the primary second-class constraint. 
	With $\gamma_a$ the first-class constraints and 
	$\chi_\alpha$ the second-class constraints, in general one has that
	\begin{equation}
		\lbrace \gamma_a , \chi_\alpha \rbrace = C_{a\alpha}^b \gamma_b 
		+ C_{a\alpha}^\beta \chi_\beta\;,
	\end{equation} 
	and the condition mentioned above is that 
	the matrix elements $C_{a\alpha_1}^{\beta \neq \beta_1}$ must be equal 
	to zero, where $\alpha_1$ and $\beta_1$ refer to the primary constraints. 
	In our case, this condition means that the quantities 
	$C_{\sigma_i \pi_1}^{\phi_1}$ should be null quantities. 
	However, 
	\begin{equation}
		\lbrace \sigma_2, \pi_1 \rbrace = \pi_1 \frac{4}{r^2} (2e_1 p_1.r + e_2 p_2.r ) + \phi_1\;,
	\end{equation} 
	showing that $C_{\sigma_2 \pi_1}^{\phi_1} \neq 0\,$. 
	Therefore, at least one condition imposed 
	in \cite{henneaux} is not satisfied, which implies that the Dirac conjecture 
	is not true in our case. 
	The first-class constraints are not necessarily all generators 
	of gauge transformations and one must use the 
	chain algorithm of \cite{Castellani} to determine these generators; see e.g. 
	Appendix \ref{AppA} for some details about it.
	%\vspace{.1cm}
	
	The field equations are obtained by taking the Poisson bracket of the
	dynamical variables with the total Hamiltonian \eqref{totalH}. 
	They explicitly read
	\begin{align}
		\dot{x}_1^\mu &= e_1 p_1^\mu + \frac{4\pi_1 }{r^2}\, e_1^2 r^\mu 
		\approx e_1 p_1^\mu \;,
		\qquad 
		\dot{x}_2^\mu = e_2 p_2^\mu +  \frac{4\pi_1}{r^2}\, e_1 e_2 r^\mu 
		\approx e_2 p_2^\mu  \;,
		\label{eomin}
		\\
		\dot{p}_1^\mu &= - \alpha^2 \frac{\sqrt{e_1 e_2}}{2r^4} r^\mu 
		+ \pi_1 \frac{8e_1}{r^4} r^\mu ( e_1 \,p_1.r + e_2 \,p_2.r ) 
		- \pi_1 \frac{4e_1}{r^2} ( e_1 \,p_1^\mu + e_2 \,p_2^\mu )
		\approx -\alpha^2 \frac{\sqrt{e_1 e_2}}{2r^4} r^\mu\;, 
		\\
		\dot{p}_2^\mu &= + \alpha^2 \frac{\sqrt{e_1 e_2}}{2r^4} r^\mu 
		- \pi_1 \frac{8e_1}{r^4} r^\mu (e_1 p_1.r + e_2 p_2.r) 
		+ \pi_1 \frac{4e_1}{r^2} (e_1 p_1^\mu + e_2 p_2^\mu )
		\approx + \alpha^2 \frac{\sqrt{e_1 e_2}}{2r^4} r^\mu\;,
		\label{EOMinit} 
		\\
		\dot{e}_1 &= \tilde{\lambda}_2\, e_1 
		-\frac{4e_1}{r^2}\, ( e_1 \,p_1 \cdot r + e_2 \, p_2 \cdot r)
		\equiv \tilde{\lambda}_2 \,e_1 + C \;,\qquad 
		\dot{e}_2 ~=~ \tilde{\lambda}_2 \,e_2 \;,
		\\
		\dot{\pi}_1 &= \phi_{1} - \tilde{\lambda}_2 \pi_1 
		+ \pi_1 ( \frac{8e_1}{r^2} p_1.r 
		+ \frac{4}{r^2} e_2 p_2.r) \approx 0 \;,
		\qquad 
		\dot{\pi}_{2} ~=~ \phi_{2} 
		- \tilde{\lambda}_2 \pi_2 + \pi_1 \frac{4}{r^2} e_1 p_2.r \approx 0 \;.
		\label{eomfin}
	\end{align} 

	The second class constraints can be dropped by adopting the Dirac bracket
	defined in terms of the inverse of the matrix
	\begin{equation} \Omega_{\alpha\beta} := \lbrace \chi_\alpha , \chi_\beta \rbrace =  
		\begin{pmatrix}
			0 & {\cal D} \\ 
			-{\cal D} & 0
		\end{pmatrix} 
		\;,
	\end{equation}
	where $(\chi_\alpha)_{\alpha=1,2} = (\pi_1,\phi_1)$ 
	denotes the two second class constraints
	and
	${\cal D} := \lbrace \pi_1 , \phi_1 \rbrace =  
	-\frac{\alpha^2}{16 r^2} 
	\sqrt{\frac{e_2}{e_1^3}}\,$. 
	One can then consider the reduced phase space 
	where one strongly sets $\pi_1$ and $\phi_1$ to 
	zero. 
	The first-class constraint $\gamma_2 \approx 0$ 
	therefore leads to the constraint $\phi_2 \approx 0\,$, 
	since the einbein $e_2$ is required to be non vanishing. 
	By using that $\phi_1$ is strongly set to zero, 
	the equation $\phi_2 \approx 0\,$ yields a constraint
	where the einbeins $e_1$ and $e_2$ have been 
	eliminated \cite{Casalbuoni:2014ofa}:
	\begin{equation}\label{masterC}
		p_1^2 p_2^2 \approx \frac{\alpha^4}{16 r^4} \;.
	\end{equation} 
	This equation is one of the main results of \cite{Casalbuoni:2014ofa}, that 
	terminates their canonical analysis.
	We pursue their analysis in order to derive a more tractable constraint 
	in view of the quantization. 
	We express \eqref{masterC} in the relative ($-$) and centre-of-mass ($+$) 
	variables 
	\begin{equation}
		p^\mu_\pm   := p_1^\mu \pm p_2^\mu\;,\qquad
		q^\mu_\pm  := \frac{1}{2} (x_1^\mu \pm x_2^\mu) \;
	\end{equation}
	that satisfy the canonical Poisson bracket relations 
	\begin{equation}
		\lbrace q_+^\mu,q_-^\nu\rbrace = 0\;,\qquad  
		\lbrace p_+^\mu,p_-^\nu \rbrace = 0\,, \qquad 
		\lbrace q_{\pm}^\mu , p_{\pm}^\nu \rbrace = \eta^{\mu \nu}\,.     
	\end{equation}
	In the new coordinate system and once the second class constraints 
	$(\chi_\alpha)_{\alpha=1,2}= (\pi_1,\phi_1)$ have been strongly set to zero, 
	the resulting first-class constraint \eqref{masterC} reads
	\begin{align}
		(p_+^2 + p_-^2 - 2p_+.p_-)(p_+^2 + p_-^2 + 2p_+.p_-) 
		- \frac{\alpha^4}{16 q_-^4 } \approx 0 \;.
		\label{FC gamma1}
	\end{align}
	In these coordinates, the canonical equations of motion 
	\eqref{eomin}-\eqref{eomfin} read:
	\begin{align}
		\dot{q}_{+}^\mu &= \frac{1}{4}( p_+^\mu (e_1 + e_2) + p_-^\mu (e_1 - e_2))\;, \label{EOMCM1}\\
		\dot{q}_{-}^\mu &= \frac{1}{4}( p_+^\mu (e_1 - e_2) + p_-^\mu (e_1 + e_2))\;, 
		\label{EOMCM2}\\
		\dot{p}_{+ \mu} &= 0 \;,\qquad 
		\dot{p}_{-}^\mu =+\frac{\alpha^2}{8 q_-^4}\, \sqrt{e_1 e_2}\, q_-^\mu \;,\\
		\dot{e}_1 &= \tilde{\lambda}_2\, e_1 + C \;,\qquad 
		\dot{e}_2 = \tilde{\lambda}_2\, e_2 \;, \qquad 
		\dot{\pi}_1 = 0 \;,\qquad \dot{\pi}_{2} = 0\;. 
		\label{EOMCM3}
	\end{align}
	Obviously, the total momentum $p_+^\mu$ is preserved by the dynamics, 
	which simply reflects the invariance of the system under constant
	spacetime translations.
	Since Eq. \eqref{FC gamma1} is not totally tractable yet, 
	we will completely fix the gauge by finding two gauge-fixing 
	conditions $C_a = 0$ such that the bracket matrix with entries 
	$M_{ab}:=\{C_a,\gamma_b\}$ is non-degenerate, effectively 
	resulting in a second-class system. 
	We propose the following gauge-fixing conditions
	\begin{align}
		C_1 := p_+ \cdot p_{-} = 0\;, \qquad 
		C_2 := e_1 e_2 - 1 = 0  \;,\label{Gauge conditions}
	\end{align}
	and readily check that the bracket matrix 
	\begin{equation}
		M =
		\begin{pmatrix}
			\{C_1, \gamma_1\} & \{C_2,\gamma_1\} \\ 
			\{C_1,\gamma_2\} & \{C_2, \gamma_2\}
		\end{pmatrix} 
		\approx
		\begin{pmatrix}
			0 & 2 e_1 e_2 \\ 
			- \alpha^2 \frac{\sqrt{e_1 e_2}}{8 q_-^4} p_+.p_- & - \frac{e_1 e_2}{q_-^2 } q_-.(p_+ (e_1+e_2) + p_-(e_1-e_2))
		\end{pmatrix} 
	\end{equation}
	is invertible, as we wished. 
	In the gauge $C_1 = 0$ 
	and in the Lorentz frame where the (preserved) total momentum 
	is $p^\mu_+ = (M,0,0,0)\,$, 
	we have
	$p^\mu_- = (0, \vec{p}_-)$ and $p_1^2 = p_2^2\,$: there is a balanced energy distribution among the two particles. Although this case will not be investigated here, it has been shown in  \cite{Sazdjian:1986km} that massless bound states can also be considered by using similar methods. 
	
	The first two field equations of \eqref{EOMCM3} indicate 
	that one can further impose the condition
	\begin{equation}
		e_1 - e_2 = 0\;,
	\end{equation} 
	which leads to $\dot{q}_+^i = 0$ and $\dot{q}_-^0 = 0\,$.
	Indeed, let us assume $e_1 = e_2\,$ and therefore $e_1 = 1 = e_2$
	on account of $C_2 = 0\,$. The two conditions $e_1 = 1 = e_2$ can 
	obviously be reached by virtue of the primary constraints \eqref{primaryC}.
	We call this gauge the \emph{unit einbein} gauge.
	Eq. \eqref{EOMCM2} shows that $q_-.p_+$ is a constant 
	that one considers to be zero in order to have $q_-^0 = 0$ in the 
	Lorentz frame adopted.
	From the first two equations \eqref{EOMCM3}, the constraint $e_1=e_2$ 
	is consistent provided 
	$C= \frac{e_1}{q_-^2}\, q_-.[ p_+ (e_1 + e_2) + p_- (e_1 - e_2)] $ 
	is zero when $e_1 - e_2 = 0\,$.
	It is straightforward to check that it is indeed the case from the 
	fact that $q_-.p_+$ is zero, as we have just shown. 
	We therefore set $q_- = (0, \vec{q}_-)\,$. 
	The three-vector $\vec{q}_+$ is set to zero, in accordance 
	with translation invariance of the system. 
	From the equations of motion and in the gauge chosen, 
	it is clear that the only dynamical variables are 
	$\vec{q}_-\,$ and $\vec{p}_-\,$. 
	
	In the gauges and Lorentz frame we have chosen, 
	Eq. \eqref{FC gamma1} becomes
	\begin{align}
		(p_+^2 + p_-^2)^2 \approx \frac{\alpha^4}{16 q_-^4 }  \;.
		\label{FC gamma1fixed}
	\end{align}
	Instead of extracting the square root of the above equation, 
	with the ambiguity in which branch to pick, we recall that 
	the remaining first-class constraint $\gamma_2\approx 0$ now reads
	$\phi_2\approx 0\,$ that leads, in the gauges we have chosen: 
	\begin{equation}
		p_+^2 + p_-^2  = - \frac{\alpha^2}{4 q_-^2 }\, \;.
	\end{equation}
	Finally one is led to the following dispersion relation 
	\begin{equation}\label{M2op}
		M^2  =\vec p^{\, 2}_- + \frac{\alpha^2}{4 \vec{q}_-^{\, 2} }\, \;.
	\end{equation}
	%
	
	%*********************************************************
	\subsection{Quantization}\label{sec:confq}
	In a Schr\"odinger quantization scheme, Eq. (\ref{M2op}) defines the eigenequation 
	\begin{equation}
		\big( - \triangle_- + \frac{\alpha^2}{4 \vec{q}_-^{~2}} \big)\ \Psi(\vec q_-) = M^2\, \Psi(\vec q_-) \;.
		\label{equation quantique}
	\end{equation}
	The spherical symmetry of 
	this operator allows to work with hyperspherical coordinates 
	$\vec q_-=({\rm q}_-,\hat \Omega_D)\,$ and to set
	\begin{equation}
		\Psi(\vec q_-) =R({\rm q}_-)\, Y_{\ell,m_a}(\hat 
		\Omega_D)\;,
	\end{equation}
	with $ Y_{\ell,m_a}(\hat \Omega_D)$ the spherical harmonics in $D$ dimensions, 
	$\ell\in\mathbb{N}$, $m_a\in\mathbb{Z}$ and $a=1,\dots , (D-1)$. Explicit forms can 
	be found for example in \cite{PhysRevA.50.3065}. More precisely, the squared mass operator 
	is a Schr\"{o}dinger Hamiltonian with repulsive inverse-squared potential:
	\begin{equation}
		- R''({\rm q}_-) -\frac{D-1}{{\rm q}_-}R'({\rm q}_-)+\frac{\ell(\ell+D-2)}{{\rm q}_-^2}R({\rm q}_-)+ \frac{\alpha^2}{4 {\rm q}_-^{\, 2}} R({\rm q}_-) = M^2\ R({\rm q}_-) \;.
		\label{equation quantique2}
	\end{equation}
	Hence the mass spectrum is continuous and the eigenstates are scattering states:
	\begin{eqnarray}\label{spec:un}
		M^2 &=& \mu^2\;,
		\nonumber \\
		R({\rm q}_-)&\sim& {({\rm q}_-)}{}^{1-\frac{D}{2}}\; J_{\lambda+\frac{D}{2}-1}(\mu\, {\rm q}_-) \;,
	\end{eqnarray}
	where $\mu\in\mathbb{R^+}$ and where $J_{\lambda+\frac{D}{2}}$ is a Bessel 
	function of the first kind. The generalized angular-momentum index $\lambda$ is 
	defined by
	\begin{equation}\label{lamdef}
		\lambda(\lambda+D-2)=\ell(\ell+D-2)+\frac{\alpha^2}{4}\;
	\end{equation}
	where $\lambda>0$ guarantees a solution that is regular at the origin. 
	Because $\alpha^2>0\,$, the radial function actually never vanishes at the origin.
	The interested reader may find in \cite{Coon:2002sua} 
	a detailed discussion of the inverse-squared potential 
	in quantum mechanics.  
	
	Our model in the conformal phase contains a continuum of bosonic states 
	with arbitrary mass $M\geq 0$. 
	We identify this continuous spectrum with unparticles, originally 
	introduced as a nontrivial scale-invariant sector in low-energy effective field 
	theories \cite{Georgi:2007ek}. As discussed in \cite{Krasnikov:2007fs, Gaete:2008aj,Nikolic:2008ax,Deshpande:2008ra}, 
	such an unparticle sector arises from a continuum of scalar fields with 
	arbitrary mass; an effective Lagrangian of the form $L\sim \partial_\mu \phi \Box^{-\delta} \partial^\mu \phi$ with $\delta >0$ is then found for a scalar unparticle \cite{Krasnikov:2007fs,Gaete:2008aj}. An originality of the present work is to provide a realization of unparticles 
	as binary states of two interacting massless particles through 
	action (\ref{actioninteraction1}). 
	
	\section{Massive particles and gapped unparticles}\label{sec:massive}
	We first recall the action for a massive relativistic particle in Minkowski spacetime:
	\begin{equation}
		S^{(m)}[x^\mu,e] = \int d\tau \; \big( \frac{\dot{x}^2}{2e}
		- \frac{e}{2} \,m^2 \big)\;,
		\label{actionpartmassive}
	\end{equation}
	where the variable $e$ is required to be nonvanishing. The above action has a smooth massless limit $m\rightarrow 0\,$.
	
	The whole constraint analysis that we have reviewed from 
	\cite{Casalbuoni:2014ofa} and presented above 
	can be repeated in the case one adds a mass term to the original action. 
	By considering the action 
	\begin{equation}\label{massiveaction}
		S_0^{(m)}[x_1^\mu,e_1,x_2^\mu,e_2] = \int d\tau \;
		\Big( \frac{\dot{x}_1^2}{2e_1} + \frac{\dot{x}_2^2}{2e_2}
		-m^2(\frac{e_1}{2}\,+\frac{e_2}{2}\,) 
		- \frac{\alpha^2}{4} \frac{\sqrt{e_1 e_2}}{r^2} \Big) \;
	\end{equation}
	that clearly breaks conformal invariance through the presence 
	of the mass terms, 
	it is straightforward to see that both the number and nature of the 
	constraints remain unchanged compared to the massless case reviewed 
	in great details above. There remains two first-class constaints 
	$\{\gamma^{(m)}_a\}\,$, $a=1,2$ 
	(one primary and one secondary) and two second-class constraints
	$\{\chi^{(m)}_\alpha\}\,$, $\alpha=1,2$
	(one primary and one secondary). 
	We adopt the same gauge-fixing conditions as in the massless case.  
	The set of conditions 
	$\{ C_1 = 0\,,~C_2 = 0\,,~\gamma^{(m)}_1 \approx 0\,,~\gamma^{(m)}_2 \approx 0\,\}$ 
	defines a set of second-class constraints so that 
	the corresponding operators each become the zero operator 
	upon quantization.
	
	In the massless case one had the constraint \eqref{FC gamma1fixed}.
	In presence of the mass terms in \eqref{massiveaction}, following the same procedure, 
	one is led to the dispersion relation
	\begin{equation}
		M^2  = \vec{p}_-^{\, 2} + \frac{\alpha^2}{4 \vec{q}_-^{\, 2}}+4m^2 \;,
		\label{M2M}
	\end{equation} 
	and, after quantization, to the spectrum (\ref{spec:un}) 
	with $M^2=\mu^2+4m^2\,$. 
	The continuous spectrum is bounded from below by  $M=2m\,$. 
	
	It has previously been noticed that coupling an unparticle to 
	an electroweak sector 
	leads to an unparticle spectrum with mass gap via a kind of Brout-Englert-Higgs 
	mechanism \cite{Delgado:2007dx}. Note also that unparticle models with a 
	mass gap are nothing but hidden-valley models \cite{Strassler:2008fv}.
	We can finally mention Ref. \cite{Megias:2019vdb}, 
	in which gapped unparticles emerge 
	as continuous Kaluza-Klein modes of a five-dimensional model with 
	a brane. 
	The action (\ref{massiveaction}) 
	may be seen as a first proposal to generate gapped unparticles from two massive 
	interacting particles. It is worth mentioning that such an action might be used in 
	the modelling of near-threshold neutral charm meson molecules, since it has been 
	argued in \cite{Braaten:2021iot} that such states (like the X(3872) which is very 
	close to the $D_0^*\bar D^0$ threshold) may be seen as unparticles.  
	%%%%
	
	\section{From the confining to the conformal phase}\label{sec:model}
	
	As starting point we propose the action 
	\begin{equation}
		S = \int~~d\tau~~ \Big[ \frac{\dot{x_1}^2}{2e_1}  + \frac{\dot{x_2}^2}{2e_2} 
		- m^2 \Big( \frac{e_1}{2}+ \frac{e_2}{2}\Big)  
		-\frac{\alpha^2 U^2(r^2)}{4} \frac{\sqrt{e_1 e_2}}{r^2} \Big].
		\label{Smodel}
	\end{equation}
	Both the mass term and the extra interaction term with the function $U(r^2)$ 
	break conformal invariance. One obviously recovers action 
	(\ref{actioninteraction1}) in the 
	limit $m\to 0$ and by setting $U = 1\,$. 
	
	The replacement of $\alpha^2$ by the position dependent coupling $\alpha^2\, U^2(r^2)$ breaks 
	conformal invariance but do not lead to  
	drastic changes in the canonical constraint analysis. 
	Although there appear terms 
	proportional to $ U'(r^2)$ from the Poisson bracket
	of the total Hamiltonian with the secondary constraints, 
	one is still able to ensure that the latter constraints 
	are preserved during dynamical evolution
	by identically solving the equation $\phi_3=0$ 
	for the function $\lambda_1\,$.
	The latter function takes a more complicated form than the 
	one given in the massless case reviewed in Section \ref{sec:review}. 
	Here we obtain
	\begin{equation}
		\lambda_1 = \frac{e_1}{e_2} \lambda_2 + C  
		+ 8\,e_1 \ln'\!U(r^2) \,
		r\cdot(p_{2} \,e_2 + p_{1}\,e_1 )\;,
	\end{equation}
	where the function $C$ is given in \eqref{lambda1}. 
	The additional terms proportional to the derivative of the function $U(r^2)$ 
	make no difference for the rest of the canonical analysis. 
	Proceeding as in the previous sections, it is straightforward 
	to find the following dispersion relation: 
	\begin{equation}\label{masterc}
		M^2=\vec p^{\, 2}_- +\frac{\alpha^2\, U^2({\rm q}_-^2)}{4{\rm q}^{\, 2}_-}
		+4m^2\;. 
	\end{equation}
	We now use the Ansatz
	\begin{equation}\label{Ua}
		U^2({\rm q}_-^2)=1+\frac{4}{\alpha^2}\sigma^2\, {\rm q}_-^4 \;,
	\end{equation}
	where the term in ${\rm q}_-^4$ can be seen as the first nontrivial term 
	in the power expansion of any function $U^2({\rm q}_-^2)$: 
	a term in ${\rm q}_-^2$ would only redefine $m$. 
	As we will show in the following, (\ref{Ua}) mimics a linear 
	confinement, typical of $(3+1)-$dimensional Yang-Mills theories in their 
	confining phase. We therefore assume that the action (\ref{Smodel}) 
	with potential (\ref{Ua}) is a relevant 
	effective model to describe binary states in gauge theories around their conformal 
	window, just as a Nambu-Goto Lagrangian is a relevant effective model for 
	light mesons ($\rho$, $f_0$,\dots) in the confined phase, see e.g. 
	\cite{Dubin:1993fk,Buisseret:2007de}. 
	
	As in Section \ref{sec:review}, we may set $q_-^0 = 0\,$, 
	so that $q_-^2 = \rm{q}_-^{\,2}\,$. With the Ansatz (\ref{Ua}), the mass operator 
	(\ref{masterc}) is now a $D-$dimensional harmonic oscillator with arbitrary 
	angular momentum: 
	$M^2=\vec p^{\, 2}_- +\frac{\alpha^2\, }{4{\rm q}^{\, 2}_-}+\sigma^2 q_-^2+4m^2\,$. Its spectrum reads \cite{PhysRevA.50.3065}
	\begin{eqnarray}
		M^2&=&2\sigma\left(2n+\lambda+\frac{D}{2}\right)+4m^2,\label{confspec}\\
		R( {\rm q}_-)&=&\left[\frac{2\Gamma(n+1)
			\sigma^{\lambda+\frac{D}{2}}}{\Gamma(n+\lambda+\frac{D}{2})} 
		\right]^{\frac{1}{2}}{\rm q}_-^{\lambda} \,
		{\rm e}^{-\frac{\sigma {\rm q}_-^2}{2}} \,
		L_n^{\lambda+\frac{D}{2}-1}(\sigma {\rm q}_-^2)\;,
	\end{eqnarray}
	with $L^\alpha_n$ the generalized Laguerre polynomials and $\lambda$ given by Eq. 
	(\ref{lamdef}). The spectrum contains massive states showing Regge trajectories, i.e., 
	$M^2\sim \ell$ or $n$ at large $\ell$ or $n$. Such a behaviour is observed 
	experimentally in light meson spectroscopy, see e.g. 
	\cite{Sonnenschein:2014jwa} and references therein. For this reason the massive 
	states we observe in the confined phase will be referred to as "mesons". 
	
An more general ansatz of the form 
		$U^2({\rm q}_-^2)=1-\frac{4}{\alpha^2}\delta^2 {\rm q}_-^2+\frac{4}{\alpha^2}\sigma^2\, {\rm q}_-^4 $ leads to the mass spectrum
		\begin{align}
			M^2=2\sigma\left(2n+\lambda+\frac{D}{2}\right)+4m^2-\delta^2\;.   
		\end{align}
		The mass scale $\delta$  could  be used to fine-tune the ground-state mass ($n=\ell=0)$. For example, the value $\delta^2=2\sigma(1+\sqrt{(D-2)^2+\alpha^2})+4m^2$ leads to a massless ground state. However, such a value causes $U^2({\rm q}_-^2)$ to be negative for some values of ${\rm q}_-^2$ since $\delta^2>\alpha\sigma$. In our model, the 
		existence of a massless ground state demands to drop the positivity of the 
		potential.
	
	Let us now focus on the low-lying confined spectrum ($n$ and $\ell$ finite) when 
	approaching the conformal window, that is in the limit $\sigma\to 0$ and $m\to 0$. According to 
	(\ref{confspec}), the mass of low-lying states will go to zero as already suggested 
	in \cite{Chivukula:1996kg,Lucini:2015noa,Lombardo:2014pda}: Light mesons are 
	expected to become massless as a signal of conformal symmetry restoration. Their 
	masses scale as $\sqrt\sigma$, which is a behaviour observed in lattice QCD in the 
	case of $SU(2)$ QCD with one adjoint Dirac quark flavour: The ratios 
	$M/\sqrt\sigma$ are found to be constant for the lightest (pseudo)scalar states 
	(mesons and glueball) as the fermion mass goes to zero in order to restore conformal 
	invariance \cite{Athenodorou:2014eua}. 
	Moreover, it is observed that the mass ratios of two 
	meson masses are constant near the conformal window as suggested by Eq. 
	(\ref{confspec}). This feature has been observed in $SU(2)$ gauge theory with two 
	adjoint quarks \cite{Bergner:2016hip}.
	
	It has to be noticed that the methodoloy used in most of the lattice QCD studies 
	of theories with conformal window is to choose $N\,$, $N_f\,$ and the quark 
	representation such 
	that the conformal window is a priori reached. The quark mass $m$ starts 
	from a nonzero value and the conformal window is reached by taking the limit 
	$m\to  0$; 
	the masses of bound states scale as $m^{\frac{1}{1+\gamma}}$ with $\gamma$ 
	the anomalous mass dimension. That parameter is then fitted on the lattice data to 
	characterise the theory under study. The interested reader may find a review of 
	computed values of $\gamma$ in Ref. \cite{Lucini:2015noa}. 
	This parameter can hardly be guessed from our effective approach: The 
	behaviours of $\sigma(m)$ and $\alpha(m)$ are not constrained by our model. 
	
	Other states of the spectrum are worth of interest: radially excited states 
	such that 
	\begin{equation}\label{limit}
		n\to\infty\quad {\rm  as}\quad \sigma\to 0\quad {\rm with}\quad \sigma\, n=\frac{\mu^2}{4}\quad \mbox{fixed}\;,
	\end{equation} 
	$\mu$ being an arbitrary (but finite) energy scale parameter.
	In this tensionless limit, the mass \eqref{confspec} remains finite.
	At large $n$ one can use a Mehler-Heine-type formula for Laguerre polynomials, see Theorem 4.1 of \cite{abramowitz+stegun}
	\begin{equation}
		L_n^{\lambda+\frac{D}{2}-1}\left(\frac{(\mu\, {\rm q}_-)^2}{4n}\right)
		\sim n^{\lambda+\frac{D}{2}-1}{\rm q}_-^{-\lambda-\frac{D}{2}+1}
		{\rm e}^{\frac{(\mu\, {\rm q}_-)^2}{8n}}\,
		J_{\lambda+\frac{D}{2}-1}(\mu\, {\rm q}_-)\; ,
	\end{equation}
	and \cite{tricomi}
	\begin{equation}
		\frac{\Gamma(n+1)}{\Gamma(n+\lambda+\frac{D}{2})}\sim n^{1-\lambda-\frac{D}{2}}.
	\end{equation}
	The spectrum (\ref{confspec}) approaches to 
	our unparticle sector (\ref{spec:un}) as $n\to+\infty$ 
	up to the rescaling $R\to \sqrt n\, R$. 
	The harmonic oscillator functions are indeed normalized to unity, which leads to the vanishing of $R$ as $\sigma\to0$ since scattering states can only be normalized to $\delta(\mu-\mu')$ in principle. 
	
	Our limit (\ref{limit}) is actually consistent with the results 
	of \cite{Stephanov:2007ry} showing that, in field theory, an unparticle 
	sector can be generated by a tower of massive states with mass $M^2_n=\Delta\ n$ 
	when the mass spacing parameter $\Delta$ goes to zero. 
	
	\section{Concluding comments}
	
\begin{figure}
		\includegraphics[width=7cm]{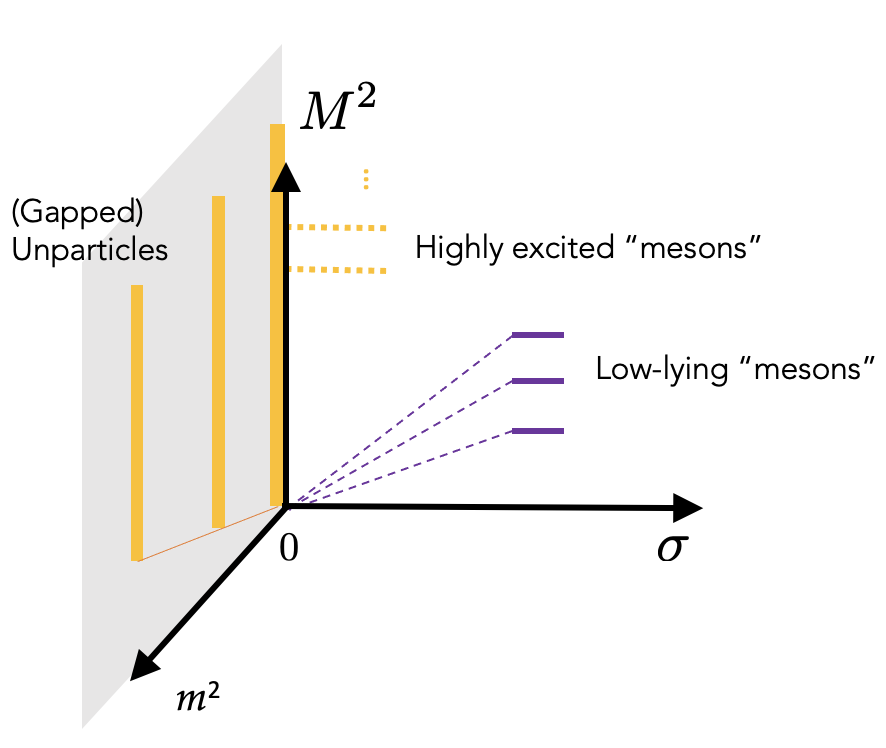}
		\caption{Schematic representation of the mass spectrum generated by the model 
			defined by Eqs. (\ref{Smodel}) and (\ref{Ua}). The behaviours of states with fixed
			$n$ (purple lines) and with fixed $\sigma n$ (dotted orange lines) are represented in the $m=0$ plane. The spectrum of gapped unparticles is represented in the $\sigma=0$ plane (solid orange lines). }\label{fig}
	\end{figure}

	The action (\ref{Smodel}) appears to be an interesting toy model to describe the 
	transition from confining to conformal phases of a Banks-Zaks-type gauge theory  
	\cite{Banks:1981nn} in terms of binary states. 
	It predicts that the low-lying bound states in the confining phase 
	become massless when approaching the conformal window with a universal behaviour, 
	the ratios of masses and $\sqrt \sigma\,$ being a constant. 
	Highly excited radial states give rise to a unparticle 
	sector in the conformal phase. The unparticle spectrum has a mass gap or not 
	depending on whether the conformal symmetry is broken or not by a mass term. A schematic drawing of this picture is given is Fig. \ref{fig}. 
	
Notice that the radial wave equation we derived in 
		the present paper also appears from a bottom-up AdS/CFT perspective, 
		see e.g. \cite{Afonin:2017fqs}.
		Interestingly, in the paper \cite{Argurio:2015wgr} where they consider a 
		$U(1)$-invariant gauge theory in AdS$_4$ propagating a vector field and 
		a complex scalar field, a nontrivial profile for the scalar field near the 
		conformal infinity of AdS$_4$ introduces two dimensionful constants 
		$m$ and $v$ that break conformal symmetry of the dual theory.
		Various limits where $m\rightarrow 0$ and $v\rightarrow 0$
		relate a discrete spectrum with a massless Goldstone boson to a 
		continuous Banks-Zaks-type spectrum for the scalar conformal operator
		in the dual CFT$_3\,$. 
		The approach of \cite{Argurio:2015wgr} enables a simultaneous discussion 
		of spontaneous and dynamical symmetry breaking in a CFT at strong coupling.
		Some bridges may presumably be drawn  between our model and the AdS/QCD framework, 
		although it is out of the scope of the present paper. 
		The interested reader may find a detailed discussions of mesons in AdS/QCD 
		in the review \cite{Erdmenger:2007cm}.

	One may finally wonder to what kind of gauge theory our toy model best applies. 
	Similarities with $SU(2)$ QCD with one adjoint quark \cite{Athenodorou:2014eua} 
	have been commented in the text. However, our model is a priori better suited to 
	model gauge theories with scalar matter than with fermionic matter, 
	the variables $x_i(\tau)$ being then identified with two "particles" of scalar matter. The existence of a conformal window is not limited to gauge theories 
	with fermionic matter, as discussed in Appendix \ref{sec:appb}. For example, a lattice study of $N_f=5$ scalar SU(2) Yang-Mills theory is affordable with 
	current computers and could be performed to check the present model. We hope that 
	such results will become available in the future. 
	
	Although we did not explicitly mention it, the present work and quoted references 
	focus on zero-temperature gauge theories. The interplay between conformal window 
	crossing and confinement/deconfinement transition may lead to new interesting 
	phenomena, as discussed in \cite{Tuominen:2012qu}. For example, a correspondence 
	has been found between conformal QCD with $N_f = 7$ and QCD with $N_f = 2$ at
	$T\approx 2\, T_c$ ($T_c$ is the deconfinement temperature). To what extent the 
	action (\ref{Smodel}) can be generalized at finite-temperature and bring relevant information is
	a problem that we leave for future works.

	\acknowledgments{We thank Zhenya Skvortsov for a discussion about 
		unparticles.}

	%\acknowledgments{In this section you can acknowledge any support given which is not covered by the author contribution or funding sections. This may include administrative and technical support, or donations in kind (e.g., materials used for experiments).}
	
\begin{appendix}
	\section{The chain algorithm for gauge generators}\label{AppA}
	
	We start from the constraint system 
	\eqref{contraintesfinales1}-\eqref{contraintesfinales2}
	and seek for the generators of gauge transformations. 
	According to the chain algorithm \cite{Castellani} 
	for the determination of the gauge generators, there is just one of 
	them because we have only one primary first-class constraint. 
	Also, the consistency algorithm stops after the secondary constraints: 
	the desired generator consists in the sum of up to two terms. 
	The constraint $\gamma_1$ itself cannot be a generator since 
	$\lbrace \gamma_1 , H_T \rbrace$ is secondary. 
	The generator is built with the first class constraints:
	\begin{equation}
		G = \dot{\varepsilon}(\tau) \gamma_1 + \varepsilon(\tau) (a \gamma_1 + b \gamma_2)\;,
	\end{equation} 
	where $a$ and $b$ are arbitrary functions at this stage. 
	Our goal is to fix these functions in order to give $G$ the properties 
	of a generator of gauge transformation. 
	Since the rules of the chain algorithm have to be respected, one imposes:
	\begin{align}
		a \gamma_1 + b \gamma_2 + \lbrace \gamma_1 , H_T \rbrace &= ~~primary\;, \\
		\lbrace a \gamma_1 + b \gamma_2 , H_T \rbrace &= ~~primary\;.
	\end{align}
	The bracket $\lbrace \gamma_1 , H_T \rbrace = \gamma_2 $ implies that $b$ must be 
	equal to $-1\,$. 
	Then, one knows that 
	$\lbrace \gamma_2 , H_T \rbrace = \tilde{\lambda}_2 \gamma_2\,$, 
	thus one concludes that the other condition implies that $a$ must be equal to $\tilde{\lambda}_2$.\\
	It can be then useful to notice the following relation:
	\begin{align}
		\dot{e}_1 = \lbrace e_1 , H_T \rbrace &= \tilde{\lambda}_2 e_1 + C \\
		\dot{e}_2 =  \lbrace e_2 , H_T \rbrace &= \tilde{\lambda}_2 e_2 \\
		&\Rightarrow ~\dot{e}_1 \,\pi_1 + \dot{e_2}\, \pi_2 
		= C \,\pi_1 + \tilde{\lambda}_2 \, \gamma_1\;.
		\label{resultatutile}
	\end{align} 
	Thanks to this relation and recalling that
	$\gamma_1 := e_1 \pi_1 + e_2 \pi_2\,$, one can rewrite 
	$G = \dot{\varepsilon} \gamma_1 + \varepsilon (\tilde{\lambda}_2 \gamma_1 
	- \gamma_2 ) $ as follows:
	\begin{equation}
		G = \frac{d}{d\tau} (\varepsilon e_1 ) \pi_1 + \frac{d}{d\tau} (\varepsilon e_2 ) \pi_2 - (\varepsilon e_1 ) \phi_1 - (\varepsilon e_2 ) \phi_2  \;.
		\label{generator}
	\end{equation}
	From this expression of the generator of gauge transformations, 
	one reads off the transformations of the variables:
	\begin{align}
		& \delta e_i = \frac{d}{d\tau} (\varepsilon e_i ) \;, 
		\qquad \delta x_i^\mu = \varepsilon \dot{x}^\mu_i\;.
		\label{transfogenerator}
	\end{align}
	These corresponds to the transformation formulae for reparametrization
	of the evolution parameter. 
	It is direct to check that the action \eqref{actioninteraction1} 
	is invariant under these transformations.

	\section{Conformal window in gauge theories with scalar matter}\label{sec:appb}
	
	The appearance of a conformal window in gauge theories with fermionic matter fields has been extensively discussed in Refs. \cite{Dietrich:2006cm} and \cite{Sannino:2009aw} for the gauge groups $SU(N)$, $SO(2N)$ and $Sp(N)$. A similar but simpler analysis can be made for a Yang-Mills theory with scalar matter 
	in the representation $R$ of the gauge algebra. The $\beta-$function of such 
	a theory is given by 
	$\beta(g)=\frac{g^3}{(4\pi)^2}\beta_0+\frac{g^5}{(4\pi)^4}\beta_1$, with $\beta_0=-\frac{11}{3}C_2({\rm adj})+\frac{N_f}{6}T(R)$, $\beta_1=-\frac{34}{3}C_2({\rm adj})^2+\frac{N_f}{3}\left(C_2({\rm adj}) 
	T(R)+6C_2(R)T(R)\right)$, 
	where $C_2({\rm adj})$ and $C_2(R)$ are the quadratic Casimir operators in the 
	adjoint and $R$ representations respectively, and where the index $T(R)$ is such 
	that $C_2(R)\dim(R)=T(R)\dim({\rm adj})$ \cite{vanDamme:1982bg}. 
	
	Applying the methodology of 
	\cite{Dietrich:2006cm} but neglecting chiral symmetry issues, one may search for a 
	conformal window in theories such that $\beta_0<0$ and $\beta_1>0$, 
	i.e., with a number of flavours such that
	\begin{equation}\label{cwind}
		\frac{34C_2({\rm adj})^2}{T(R)\left(C_2({\rm adj})+6 C_2(R)\right)}< N_f 
		<\frac{22 C_2({\rm adj})}{T(R)},
	\end{equation}
	the coupling constant $g^*$ at which it is observed being equal to 
	$g^{*2}=-\frac{16\pi^2\beta_0}{\beta_1}$. The equation (\ref{cwind}) admits 
	nontrivial solutions. A simple example is the case of matter in adjoint 
	representation, for which a conformal phase appears if $5<N_f<22$. 
	
	\end{appendix}
	
	%\reftitle{References}
	

\end{document}